\documentclass[12pt]{article}      
\usepackage{epsfig,multirow,setspace}

\AtEndOfClass{\RequirePackage{times}}
\usepackage{mathtools}
\usepackage{amsthm,amsmath,amssymb,colonequals,multirow,graphicx,epstopdf}
\usepackage{caption}
\usepackage{color}
\usepackage{mwe}

\usepackage[caption=false,font=footnotesize,labelfont=sf,textfont=sf]{subfig}
\usepackage{fancyhdr}
\usepackage{amsfonts, amsmath, amssymb, marvosym, upgreek, mathrsfs, dsfont} 
\usepackage{array,graphicx}
\usepackage{booktabs}
\usepackage{pifont}
\usepackage{setspace}
\usepackage{colonequals} 
\usepackage[authoryear]{natbib}

\renewcommand\harvardyearright[1]{.}
\newcommand\Fontvi{\fontsize{8}{7.2}\selectfont}

\newcommand{\inv}{^{\raisebox{.2ex}{$\scriptscriptstyle-1$}}}
\DeclareMathOperator{\E}{\mathbb{E}}
\DeclareMathOperator{\R}{\mathbb{R}}

\begin{document}

\title{A Multivariate Poisson-Log Normal Mixture Model for Clustering Transcriptome Sequencing Data}

\author{Anjali Silva\footnote{Department of Mathematics and Statistics, University of Guelph, Guelph, ON, Canada.}\qquad\qquad\qquad Steven~J.~Rothstein\footnote{Department of Molecular and Cellular Biology, University of Guelph, Guelph, ON, Canada.}\\\hspace{-0.4in}Paul~D.~McNicholas\footnote{Department of Mathematics and Statistics, McMaster University, Hamilton, ON, Canada.} \qquad\qquad Sanjeena Subedi\footnote{Department of Mathematical Sciences, Binghamton University, Binghamton, New York, USA.}}
\date{}

\maketitle

\begin{abstract}
High-dimensional data of discrete and skewed nature is commonly encountered in high-throughput sequencing studies. Analyzing the network itself or the interplay between genes in this type of data continues to present many challenges. As data visualization techniques become cumbersome for higher dimensions and unconvincing when there is no clear separation between homogeneous subgroups within the data, cluster analysis provides an intuitive alternative. The aim of applying mixture model-based clustering in this context is to discover groups of co-expressed genes, which can shed light on biological functions and pathways of gene products. A mixture of multivariate Poisson-Log Normal (MPLN) model is proposed for clustering of high-throughput transcriptome sequencing data. The MPLN model is able to fit a wide range of correlation and overdispersion situations, and is ideal for modeling multivariate count data from RNA sequencing studies. Parameter estimation is carried out via a Markov chain Monte Carlo expectation-maximization algorithm (MCMC-EM), and information criteria are used for model selection.
\end{abstract}

\section{Introduction}\label{sec:intro}
With the advent of genomics studies, it is now possible to look at the genome at a global scale. An integral component of genomics studies is functional genomics, which involves studying the expression and interaction of genes. Expression of genes can be identified using RNA transcripts and the process of sequencing RNA transcripts is called RNA sequencing (RNA-seq). RNA-seq can be used to determine the transcriptional dynamics of a biological system by measuring the expression levels of thousands of genes simultaneously \citep{wang2009, roberts2011}. This technique provides counts of reads that can be mapped back to a biological entity, such as a gene or an exon, which is a measure of the gene's expression under experimental conditions. Analyzing RNA-seq data is challenged by several factors, including the nature of the data, which is characterized by high dimensionality, skewness, and presence of a dynamic range that may vary from zero to over a million counts. Further, multivariate count data from RNA-seq is generally overdispersed. Upon obtaining raw counts of reads from an RNA-seq study, a typical bioinformatics analysis pipeline involves trimming, mapping, summarizing, normalizing and downstream analysis \citep{oshlack2010}. Cluster analysis is often performed as part of downstream analysis to identify key features between observations. 

Clustering algorithms can be classified into two broad categories: distance-based or model-based approaches \citep{zhong2003}. Distance-based clustering techniques include hierarchical clustering and partitional clustering \citep{zhong2003}. Distance-based approaches utilize a distance function between pairs of data objects and group similar objects together into clusters. However, discriminative approaches can be computationally inefficient requiring a complexity of $\mathcal{O}(N^2)$ as the approach involves comparing all pairs of data objects, $N$ \citep{zhong2003}. Further, it involves non-fuzzy or hard clustering, where each data point can only belong to one cluster. Model-based approaches involve clustering data objects using a mixture-modeling framework \citep{wolfe1965, mclachlanandbasford1998, mclachlan2000, zhong2003, mcnicholas2016}. Compared to discriminative approaches, model-based approaches offer better interpretability because the resulting model for each cluster directly characterizes that cluster \citep{zhong2003}. In model-based approaches, hard clustering is not required as the conditional probability of each data object belonging to a cluster is calculated. 

The general distribution function of a mixture model can be given as
\begin{equation}
\label{eqn:gdis}
f( \boldsymbol{y}| \pi_1, \ldots, \pi_G,\boldsymbol{\vartheta}_1, \ldots, \boldsymbol{\vartheta}_G ) = \sum_{g=1}^{G} \pi_g f_g( \boldsymbol{y} | \boldsymbol{\vartheta}_g),
\end{equation}
where $G$ is the total number of clusters, $f_g(\cdot)$ is the distribution function with parameters $\boldsymbol{\vartheta}_g$, and $ \pi_g>0$ is the mixing weight of the $g^{\text{th}}$ component such that $\sum_{g=1}^G \pi_g=1$. As the cluster membership of all observations is assumed to be unknown, an indicator variable $z_{ig}$  is used for cluster membership, such that ${z}_{ig}$ equals 1 if the $i$th observation belongs to component $g$ and $0$ otherwise. The predicted group memberships at the maximum likelihood estimates of the model parameters are given by the maximum \textit{a posteriori} probability, $\text{MAP}(\hat{z}_{ig}$). The $\text{MAP}(\hat{z}_{ig}) = 1$ if $\text{max}_h(\hat{z}_{ih}), h = 1, \ldots, G$, occurs at component $g$ and $\text{MAP}(\hat{z}_{ig}) = 0$ otherwise. The complete-data consist of the predicted group memberships, the mixing weights, and the observed data. Parameters and group membership estimation is carried out using the maximum likelihood algorithms, such as the expectation-maximization (EM) algorithm \citep{dempster1977}. As the number of clusters in the finite mixture model is almost always unknown, the parameter estimation methods are fitted for a range of possible number of components and the optimal number is selected using a model selection criterion. Typically, one component represents one cluster \citep{mcnicholas2016}. 

\section{Clustering RNA-seq data}
Clustering of RNA-seq data allows identifying groups of genes with similar expression patterns, called gene co-expression networks. Inference of gene networks from RNA-seq data can lead to better understanding of biological pathways and regulatory mechanisms that are active under experimental conditions. This information can also be used to infer the biological function of genes with unknown or hypothetical functions based on their cluster membership with genes of known functions and pathways \citep{dhaeseleer2005}. Over the past few years, a number of mixture model-based clustering approaches for RNA-seq data have emerged based on the univariate Poisson and negative binomial (NB) distributions \citep{rau2011, papastamoulis2014, Si2014}. Although, these distributions seem a natural fit to count data, there can be limitations when applied in the context of RNA-seq as outlined in the following paragraph.

The Poisson distribution is used to model discrete data, including expression data from RNA-seq studies. However, multivariate extension of the Poisson distribution can be computationally expensive. As a result, the univariate Poisson distribution is often utilized in clustering algorithms, which leads to the assumption that samples are independent conditionally on the components \citep{rau2011, papastamoulis2014, rau2015}. This assumption is unlikely to hold in real situations. Further, the mean and variance coincide in the Poisson distribution. As a result, the Poisson distribution may provide a good fit to RNA-seq studies with a single biological replicate across technical replicates \citep{kvam2012}. However, current RNA-seq studies often utilize more than one biological replicate in order to estimate the biological variation between treatment groups. In such studies, RNA-seq data exhibit more variability than expected (called ``overdispersion") and the Poisson distribution may not provide a good fit for the data \citep{kvam2012, anders2010}. Due to the smaller variation predicted by Poisson distribution, type-I errors in the data can be underestimated 
\citep{anders2010}. The use of NB distribution may alleviate some of these issues as the mean and variance differ. However, NB can fail to provide a good fit to heavy tailed data like RNA-seq \citep{esnaola2013}.

The MPLN distribution \citep{aitchison1989} is a multivariate log normal mixture of independent Poisson distributions. It is a two-layer hierarchical model, where the observed layer is a multivariate Poisson distribution and the hidden layer is a multivariate Gaussian distribution \citep{aitchison1989, vera2011}. The MPLN distribution is suitable for analyzing multivariate count measurements and offers many advantages over other discrete distributions. Importantly, the hidden layer of the MPLN distribution is a multivariate Gaussian distribution, which allows for the specification of a covariance structure. As a result, independence no longer needs to be assumed between variables. For the MPLN distribution, the $\mathds{V}\text{ar}(Y_j) \geq \E(Y_j)$, so there is overdispersion for the marginal distribution with respect to the Poisson distribution. Thus, the MPLN distribution can account for overdispersion in count data. Further, this distribution supports negative and positive correlations, unlike other multivariate discrete distributions such as multinomial or negative multinomial \citep{tunaru2002}.

Here, we present a novel mixture model-based clustering method for RNA-seq using the MPLN distribution. We explore the proposed clustering technique in the context of clustering genes. All analyses were performed on a MacBook Pro with 3.1 GHz quad-core Intel Core i7 processor and 16 GB RAM. The code was developed using \textsf{R} and Rstudio \citep{r2014, rstudio2012}. 

\section{Methods}\label{sec:method}
\subsection{Multivariate Poisson-Log Normal Mixture}
For samples $i \in \{1, \ldots, n\}$ and dimensionality $j \in \{1, \ldots,d\}$, the MPLN is defined via:
\begin{equation}\begin{split}
\label{eqn:MPLN}
                  Y_{ij} | \theta_{ij} &\sim \mathscr{P}(e^{\theta_{ij}})\\
                  (\theta_{i1}, \ldots,\theta_{id})^{\prime} &\sim \mathscr{N}_{d}(\boldsymbol{\mu},\boldsymbol{\Sigma}),
\end{split}
\end{equation}
where $\mathscr{N}_d(\boldsymbol{\mu},\boldsymbol{\Sigma})$ is the $d$-dimensional normal distribution with mean vector $\boldsymbol{\mu}$ and covariance matrix $\boldsymbol{\Sigma}$. The $\mathscr{P}$ denotes Poisson distribution with parameters $e^{\theta_{ij}}$. The $Y_{ij}$ represents the observed count and $\theta_{ij}$ represents the latent variable of the model. The moments of the MPLN distribution can be obtained via conditional expectation results and standard properties of the Poisson and log normal distributions \citep{aitchison1989}. Because the sequencing depth can differ between samples involved in an RNA-seq study, the assumption of equal means across conditions is unlikely to hold. To account for the differences in library sizes, a fixed, known constant, $s_{j}$, representing the normalized library sizes are added to the mean of the Poisson distribution. Thus, for genes $i \in \{1, \ldots, n\}$ and samples $j \in \{1, \ldots,d\}$, the MPLN distribution is modified to give:
\begin{equation}\begin{split}
\label{eqn:MPLN2}
                  Y_{ij} | \theta_{ij} &\sim \mathscr{P}(e^{\theta_{ij}+{\log(s_{j})}} )\\
                  (\theta_{i1}, \ldots,\theta_{id})^{\prime} &\sim \mathscr{N}_{d}(\boldsymbol{\mu},\boldsymbol{\Sigma}).
\end{split}
\end{equation}
 A $G$-component mixture of MPLN distributions can be written as
\vspace{-0.1in} 
\[
 f(\mathbf{y};\boldsymbol{\Theta})=\sum_{g=1}^G\pi_gf_{\mathbf{Y}}(\mathbf{y}|\boldsymbol{\mu}_g,\boldsymbol{\Sigma}_g) = \sum_{g=1}^G\pi_g\int_{\R^d} \left( \prod _{j=1}^d f(y_{ij}| \theta_{ijg},  s_j)\right) ~f(\boldsymbol{\theta}_{ig}|\boldsymbol{\mu}_g,\boldsymbol{\Sigma}_g)~d\boldsymbol{\theta}_{ig},
\]
where $\boldsymbol{\Theta}$ denotes all model parameters and $f_{\mathbf{Y}}(\mathbf{y};\boldsymbol{\mu}_g,\boldsymbol{\Sigma}_g)$ denotes the distribution of the $g^{th}$ component with parameters $\boldsymbol{\mu}_g$ and $\boldsymbol{\Sigma}_g$. In the context of clustering, the unknown component membership indicator variable is denoted by $\mathbf{Z}$ such that $Z_{ig}=1$ if an observation $i$ belongs to group $g$ and $Z_{ig}=0$ otherwise.

\subsection{Parameter estimation}
To estimate the parameters, a maximum likelihood estimation (MLE) procedure based on EM algorithm is utilized. EM is a popular method for computing MLE in problems with missing data. In the expectation (E) step, the conditional expectation of complete-data log-likelihood  given observed data $\textbf{y}$ and current model parameters are calculated. During the maximization (M) step the expected value of complete-data log-likelihood is maximized with respect to the model parameters. The E and M-steps are iterated until some convergence criterion is met. The complete-data log-likelihood for the MPLN mixture model is
\begin{equation*}
\label{eqn:clogL}
\begin{split}
l_{c}(\boldsymbol{\pi}, \boldsymbol{\vartheta} | \boldsymbol{y},& \boldsymbol{\theta}, \boldsymbol{s}, \boldsymbol{z}) = \sum_{i=1}^{n} \sum_{g=1}^G z_{ig} \log \pi_g \Big( \sum _{j=1}^d f(y_{ij}| \theta_{ijg}, s_j)\Big) ~f(\boldsymbol{\theta}_{ig}|\boldsymbol{\mu}_g,\boldsymbol{\Sigma}_g)\\
 =& \sum_{g=1}^G  n_g \log \pi_g - \sum_{i=1}^n \sum_{g=1}^G \sum_{j=1}^d z_{ig} \exp \{\theta_{ijg} + \log s_j\} + \sum_{i=1}^n \sum_{i=g}^G z_{ig}  (\boldsymbol{\theta}_{ig} + \log \mathbf{s}) \mathbf{y}_i^{\prime} \\
& -  \sum_{i=1}^n \sum_{g=1}^G \sum_{j=1}^d z_{ig} \log \mathbf{y}_{ij} ! - \frac{nd}{2} \log 2 \pi - \frac{1}{2} \sum_{g=1}^G n_g \log | \mathbf{\Sigma}_g |\\ 
&- \frac{1}{2} \sum_{i=1}^n \sum_{g=1}^G z_{ig} (\boldsymbol{\theta}_{ig} - \boldsymbol{\mu}_g) \mathbf{\Sigma}_g^{\inv} (\boldsymbol{\theta}_{ig} - \boldsymbol{\mu}_g)^{\prime},
\end{split}
\end{equation*}
where $\boldsymbol{\vartheta} = (\boldsymbol{\mu}_g, \boldsymbol{\Sigma}_g)$ and $n_g =  \sum_{i=1}^n z_{ig}^{(t)}$. 
The conditional expectation of complete-data log-likelihood given observed data ($\mathcal{Q}$) is 
\begin{equation}
\label{eqn:Q}
\begin{split}
\mathcal{Q}(\boldsymbol{\pi}, \boldsymbol{\vartheta}| \boldsymbol{\pi}^{(t)}, \boldsymbol{\vartheta}^{(t)}) & = \mathds{E}_{\vartheta^{(t)}} \big\{l_c(\boldsymbol{\pi}, \boldsymbol{\vartheta} | \boldsymbol{y}, \boldsymbol{\theta}, \boldsymbol{s}, \boldsymbol{z}) \big\}
 = \mathds{E}_{\vartheta^{(t)}} \big[ \log \big(\pi_g f(\boldsymbol{y}|\boldsymbol{\theta}, \boldsymbol{s}) f(\boldsymbol{\theta} | \boldsymbol{\vartheta}) \big) \big].
\end{split}
\end{equation}
Because the first two terms of \eqref{eqn:Q} does not depend on parameters $\boldsymbol{\vartheta} $, the $\mathcal{Q}$ can be written
\begin{equation}
\label{eqn:newQ}
\begin{split}
\mathcal{Q}(\boldsymbol{\vartheta}| \boldsymbol{\vartheta}^{(t)}) & = \mathds{E}_{\vartheta^{(t)}} \big[\log f(\boldsymbol{\theta} | \boldsymbol{\vartheta}) | \boldsymbol{Y} = \boldsymbol{y} \big] + c(\boldsymbol{y}).
\end{split}
\end{equation}
The density of the term $ f(\boldsymbol{\theta}| \boldsymbol{y}, \boldsymbol{\vartheta})$ in (\ref{eqn:newQ}) is
\begin{equation}
\label{eqn:expectation}
\begin{split}
f(\boldsymbol{\theta} | \boldsymbol{y}, \boldsymbol{\vartheta}) & = \frac{f(\boldsymbol{y} | \boldsymbol{\theta}) f(\boldsymbol{\theta}, \boldsymbol{\vartheta})} {f(\boldsymbol{y}, \boldsymbol{\vartheta})} = \frac{f(\boldsymbol{y} | \boldsymbol{\theta}) f(\boldsymbol{\theta}, \boldsymbol{\vartheta})} {\int f(\boldsymbol{y} | \boldsymbol{\theta}) f(\boldsymbol{\theta}, \boldsymbol{\vartheta})  \partial \boldsymbol{\theta} }.
\end{split}
\end{equation}
Due to the integral present in (\ref{eqn:expectation}), evaluation of $f(\boldsymbol{y}|\boldsymbol{\vartheta})$ is difficult. Therefore, the E-step cannot be solved analytically. Here, an extension of the EM algorithm, called Monte Carlo EM (MCEM), can be used to approximate the $\mathcal{Q}$ function \citep{wei1990a}. The Monte Carlo EM involves simulating at each iteration $t$ and for each observation $\boldsymbol{y}_i$ a random sample $\boldsymbol{\theta}^{(t)}_{i1}, \ldots,\boldsymbol{\theta}^{(t)}_{iN}$ from the distribution $f(\boldsymbol{\theta}|\boldsymbol{y},\boldsymbol{\vartheta})$ to find a Monte Carlo approximation to the conditional expectation of complete data log-likelihood given observed data. Thus, a Monte Carlo approximation for $\mathcal{Q}$ in (\ref{eqn:newQ}) is
\begin{equation*}
\begin{split}
\mathcal{Q}(\boldsymbol{\vartheta}| \boldsymbol{\vartheta}^{(t)}) & =\sum_{g=1}^G \sum_{i=1}^nQ_{ig}(\boldsymbol{\vartheta}| \boldsymbol{\vartheta}^{(t)}),\\
\mathcal{Q}_{ig}(\boldsymbol{\vartheta}| \boldsymbol{\vartheta}^{(t)}) & \simeq \frac{1}{N} \sum_{k=1}^N \log f(\boldsymbol{\theta}_{ig}^{(k)}|\boldsymbol{y}_i,\boldsymbol{\vartheta}) + c(\boldsymbol{y}_i).
\end{split}
\end{equation*}
However, another layer of complexity is added as the distribution of $f(\boldsymbol{\theta}|\boldsymbol{y},\boldsymbol{\vartheta})$ is unknown. Therefore, an alternative MCEM based on Markov chains (MCMC-EM) via Stan is proposed. The \textsf{R} version of Stan is available via ${\tt RStan}$ \citep{rstan}.
\subsection{Bayesian inference with Stan}
Bayesian approaches to mixture modeling offer the flexibility of sampling from computationally complex models via Markov chain Monte Carlo (MCMC) sampling algorithms. MCMC algorithms make use of Markov chains to approximate the posterior distribution, given the prior distribution and likelihood. MCMC is ideal when a closed-form solution to posterior distribution is not available. The aim of MCMC is to create a Markov process that has a stationary distribution the same as the posterior distribution. Stan is a software package for MCMC applications that utilizes No-U-Turn Sampler (NUTS) \citep{stan}. NUTS is an extension of the Hamiltonian Monte Carlo (HMC) MCMC algorithm and is able to sample from posterior distributions with correlated parameters \citep{nuts2014}. Further, NUTS requires no tuning parameters \citep{nuts2014}.

For the mixtures of MPLN distribution, the random sample $\boldsymbol{\theta}^{(t)}_{i1}, \ldots,\boldsymbol{\theta}^{(t)}_{iN}$ is simulated via ${\tt RStan}$ package (version 2.15.1). The prior on $\boldsymbol{\theta}$ is a multivariate Gaussian distribution and the likelihood follows a Poisson distribution. Within ${\tt RStan}$, the ${\tt warmup}$ argument is set to half the number of total iterations, as recommended \citep{stan}. The warmup samples are used to tune the sampler in which Stan optimizes the HMC algorithm and are discarded from further analysis. 

\subsection{Parameter estimation}
Using MCMC-EM, the expected value of $\boldsymbol{\theta}$ and group membership variable $z_{ig}$, respectively, are updated in E-step as follows:
\begin{equation*}\begin{split}
&\E_{\vartheta^{(t)}}( \boldsymbol{\theta}_{ig} |\textbf{y}_i) \simeq \frac{1}{N} \sum_{k=1}^N \boldsymbol{\theta}_{ig}^{(k)} \simeq \boldsymbol{\theta}^{(t)}_{ig},\\
&\E_{\vartheta^{(t)}} (z_{ig} |\textbf{y}_i, \boldsymbol{\theta}_i, \mathbf{s}) = \frac{ \pi_g f(\textbf{y}_{i}| \boldsymbol{\theta}_{ig}^{(t)}, \mathbf{s})  f(\boldsymbol{\theta}_{ig}|\boldsymbol{\mu}_g^{(t)}, \boldsymbol{\Sigma}_g^{(t)})}{\sum_{h=1}^G \pi_h^{(t)} f(\boldsymbol{y}_{i}|\boldsymbol{\theta}_{ih}^{(t)}, \mathbf{s})  f(\boldsymbol{\theta}_{ih}|\boldsymbol{\mu}_h^{(t)}, \boldsymbol{\Sigma}_h^{(t)})} \equalscolon z_{ig}^{(t)}.
\end{split}\end{equation*} 
M-step has an explicit solution, thus maximum likelihood estimate of the parameters are obtained as follows
\begin{equation*}\begin{split}
&\pi_g^{(t+1)} = \frac{\sum	_{i=1}^n z_{ig}^{(t)}}{n},\qquad
\boldsymbol{\mu}^{(t+1)}_{g} = \frac{\sum _{i=1}^n z_{ig}^{(t)} \E (\boldsymbol{\theta}_{ig})} {\sum_{i=1}^n z_{ig}^{(t)}},\\
&\small \mathbf{\Sigma}^{(t+1)}_{g} = \frac{\sum_{i=1}^n z_{ig}^{(t)} \E \Big((\boldsymbol{\theta}_{ig} - \boldsymbol{\mu}^{(t+1)}_g) (\boldsymbol{\theta}_{ig} - \boldsymbol{\mu}^{(t+1)}_g)^{\prime} \Big)}  {\sum_{i=1}^n z_{ig}^{(t)} }.
\end{split}\end{equation*}

\subsection{Convergence}
To determine whether the MCMC chains have converged to the posterior distribution, several diagnostic criteria are used. One criteria is the potential scale reduction factor, $\hat{R}$ \citep{gelman1992}. $\hat{R}$ gives a measure of between-chain variance relative to that of within-chain variance. A large $\hat{R}$ indicates higher between-chain variance relative to within-chain variance, thus need for longer simulation. If $\hat{R}$ is close to 1, it can be concluded that each of the chains have reached the target distribution. Another criteria is the effective number of samples, ${N_{\text{\tiny eff}}}$ \citep{gelman2013}, which provides the number of independent samples represented in the chain. As recommended, the algorithm for mixtures of MPLN distribution is set to check if the ${\tt RStan}$ generated chains have a $\hat{R}$ less than 1.1 and an ${N_{\text{\tiny eff}}}$ value greater than 100 \citep{stan}. If both criteria are met, the algorithm proceeds. Otherwise, the chain length is set to increase by 100 iterations and sampling is redone. A total of $3$ chains are run at once, as recommended \citep{stan}. 

Despite the wide popularity of MCMC for various applications, difficulties can be encountered when determining convergence of MCMC chains. It is recommended that automated convergence check is unsafe and should be avoided for MCMC \citep{cowles1996}. Further, it is not possible to guarantee that a finite sample from an MCMC is representative of the underlying stationary distribution \citep{cowles1996}. However, convergence diagnostics such as $\hat{R}$ and ${N_{\text{\tiny eff}}}$, along with traceplots, should provide an overview of the algorithm's progress. It is recommended that Monte Carlo sample size should be increased with the MCMC-EM iteration count due to persistent Monte Carlo error \citep{neath2012}, which can contribute to slow or no convergence. For the algorithm of mixtures of MPLN distribution, the number of ${\tt RStan}$ iterations is set to start with a modest number of 1000 and is increased with each MCMC-EM iteration as the algorithm proceeds. 

The MCMC-EM lacks the ascent property as the MCEM update maximizes an estimate of the $\mathcal{Q}$ rather than $\mathcal{Q}$ itself \citep{neath2012}. Therefore, convergence criteria based on the ascent property can run into the risk of terminating early due to Monte Carlo error \citep{hobert1999, neath2012}. We utilized the Heidelberger and Welch's convergence diagnostic, which uses the Cramer-von-Mises statistic to test the null hypothesis that the sampled values come from a stationary distribution \citep{heidelberger1983}. The diagnostic is implemented via the ${\tt heidel.diag}$ function in ${\tt coda}$ package (version 0.19-1) \citep{coda}. The test is applied to all log-likelihood values up to the current MCMC-EM iteration using a significance level of 0.05.
\subsection{Initialization}
For initialization of parameters $\boldsymbol{\mu}$ and $\boldsymbol{\Sigma}$, the $\log$ of the mean and covariance of each dataset is obtained via the ${\tt mean}$ and ${\tt cov}$ functions in \textsf{R} are used, respectively. For initialization of $\hat{z}_{ig}$, two algorithms are provided: $k$-means and random. For $k$-means initialization, $k$-means clustering is performed on the dataset and the resulting cluster memberships are used for the initialization of $\hat{z}_{ig}$. The algorithm is then run for 10 iterations and the resulting $\hat{z}_{ig}$ values are used as starting values. For random initialization, random values are chosen for $\hat{z}_{ig} \in \{0,1\}$ such that $\sum_{i=1}^n \hat{z}_{ig} = 1$ for all $i$. The algorithm is then run for 10 iterations and resulting $\hat{z}_{ig}$ values are used as starting values. If multiple initialization runs are considered, the $\hat{z}_{ig}$ values corresponding to the run with the highest log-likelihood value are used for downstream analysis.
\subsection{Parallel implementation}
Coarse grain parallelization has been developed in the context of model-based clustering of Gaussian mixtures \citep{mcnicholas2010parallel}. When a range of clusters are considered for a dataset, i.e., $G_{\text{min}}$:$G_{\text{max}}$, each cluster size, $G$, is independent and have no dependency between them. Therefore, each $G$ can be run in parallel, each one on a different processor. Here, the algorithm for mixtures of MPLN distribution was parallelized using ${\tt parallel}$ package \citep{r2014} and ${\tt foreach}$ package (version 1.4.3) \citep{foreach2015}. Parallelization reduced the running time of the datasets (results not shown). All analyses of mixtures of MPLN distribution were done using the parallelized code. 
\subsection{Model selection}
Typically, a dataset is run for a range of possible cluster sizes and the best among these is selected using some criterion. As the log-likelihood value favours the model with more parameters, model selection criteria are used that penalize the log-likelihood for model complexity. Among the different criteria available, the Bayesian information criterion \citep[BIC;][]{schwarz1978} remains as the most popular criterion for model-based clustering applications \citep{mcnicholas2016}. For this analysis, we utilized four model selection criteria: the Akaike information criterion \citep{akaike1973}, $$\text{AIC} = -2 \log \mathcal{L} (\hat{\boldsymbol{\vartheta}} |\boldsymbol{y}) + 2K;$$ the BIC, $$\text{BIC} = -2 \log \mathcal{L} (\hat{\boldsymbol{\vartheta}} |\boldsymbol{y}) + K \log (n);$$ a variation on the AIC used by \cite{Bozdogan1994}, $$\text{AIC3} =  -2 \log \mathcal{L} (\hat{\boldsymbol{\vartheta}}|\boldsymbol{y}) + 3K;$$ and the integrated completed likelihood (ICL) of \cite{biernacki2000}, $$\text{ICL} \approx \text{BIC} + 2 \sum_{i=1}^n \sum_{g=1}^G \text{MAP}\{ \hat{z}_{ig}\} \log \hat{z}_{ig}.$$ The $\mathcal{L} (\hat{\boldsymbol{\vartheta}} |\mathbf{y})$ represents maximized log-likelihood, $\hat{\boldsymbol{\vartheta}}$ is the maximum likelihood estimate of the model parameters $\boldsymbol{\vartheta}$, $n$ is the number of observations, and $\text{MAP}\{ \hat{z}_{ig}\}$ is the maximum $\textit{a posteriori}$ classification given $\hat{z}_{ig}$. $K$ represents the number of free parameters in the model, calculated as $K = (G-1) + (Gd) + Gd(d+1)/2$, for $G$ clusters. 

These model selection criteria differ in terms of how they penalize the log-likelihood. The AIC and AIC3 penalize log-likelihood only for the number of free parameters in the model and are constant with respect to the sample size. And hence, when $n$ is large, AIC tends tends to favor more complex models \citep{shibata1976,katz1981} and it overestimates number of components in the model.  BIC penalizes the log-likelihood based on both the number of  free parameters in the model and the number of observations. The BIC selects the number of mixture components needed to provide a good approximation to the density, rather than the number of clusters. As a result, BIC often assigns multiple mixture components to a single cluster. ICL penalizes BIC for the estimated mean entropy based on the spread of the mixture components. The ICL favors well separated clusters compared to BIC \citep{biernacki2000}.

\subsection{Performance assessment}
Performance assessment with respect to clustering was done using adjusted Rand index \citep[ARI;][]{hubert1985}, which has become the most popular assessing method in mixture model-based applications \citep{mcnicholas2016}. The ARI measures pair wise agreement between the predicted and true cluster memberships. ARI has an expected value of zero under random classification and a value of one for perfect agreement. Negative values of ARI may result when the classifications are worst than that would be expected under random classification. 

\subsection{Simulation study}
We considered different cluster sizes such that $50$ datasets with two underlying clusters and $30$ datasets with three underlying clusters were generated. Simulated datasets with $n = 1000$ observations and $d = 6$ samples, were generated using mixtures of MPLN distribution. Count range in datasets represented 5--95\% count range observed in real RNA-seq dataset (see Table~\ref{table_realdata}). The covariance matrices for each setting is generated using the ${\tt genPositiveDefMat}$ function in ${\tt clusterGeneration}$ package (version 1.3.4) \citep{clusterGeneration2015}. Initialization of $z_{ig}$ is done using $k$-means algorithm with $3$ runs, and each run consists of 10 iterations. For the two clusters setting, $\pi_1 = 0.79$ and a clustering range of $G = 1, \ldots , 3$ was considered. For the three clusters setting, $\pi_1 = 0.3$ and $\pi_2 = 0.5$, and a clustering range of $G = 2, \ldots , 4$ was considered.
\subsection{Comparative study}
Three methods utilized for this study include ${\tt HTSCluster}$ package (version 2.0.4) \citep{rau2011, rau2015}, ${\tt MBCluster.Seq}$ package (version 1.0) \citep{Si2014}, and ${\tt Poisson.glm.mix}$ package (version 1.2) \citep{papastamoulis2014}. For details on the methods, we refer the reader to original articles. To make settings as similar as possible across all methods, normalization factors representing library size estimate for each sample is obtained using trimmed mean of M values (TMM) method from ${\tt calcNormFactors}$ function of ${\tt edgeR}$ package (version 3.17.10) \citep{edgeR2010, robinson2010, mccarthy2012}. Initialization is done via $k$-means for ${\tt HTSCluster}$ and ${\tt MBCluster.Seq}$. An option to specify normalization or initialization method was not available for ${\tt Poisson.glm.mix}$, thus default settings were used. For all methods, three initialization runs, each consisting of $10$ iterations (or less iterations if converged), are conducted prior to clustering via EM algorithm. The methods were applied to the same datasets generated in the simulation study with two and three underlying clusters. For the two clusters setting, a clustering range of $G = 1, \ldots , 3$ was considered and for three clusters setting, a clustering range of $G = 2, \ldots , 4$ was considered. Note, for ${\tt MBCluster.Seq}$, $G = 1$ cannot be run. Therefore clustering was run for a range of $G = 2, \ldots , 3$ for the two clusters setting only. 
\subsection{Transcriptome data analysis}
Raw read counts for 1336 genes, which were identified as differentially expressed by Freixas Coutin et al. were obtained from Binary Alignment/Map (BAM) files using samtools (version 0.1.17) \citep{li2009}  and HTSeq (version 0.6.1p2) \citep{anders2015}. The median value from the $3$ replicates per each developmental stage was chosen and data was clustered for a range of $G = 1, \ldots , 11$ with three $k$-means initialization runs. Note, for ${\tt MBCluster.Seq}$, $G = 1$ cannot be run. Therefore clustering was run for a range of $G = 2, \ldots , 11$. In the context of real data clustering, it is not possible to compare the clustering results obtained from each method to a `true' clustering of the data as such classification does not exist. To identify if co-expressed genes are implicated in similar biological processes or pathways, a gene ontology (GO) enrichment analysis was performed on the gene clusters. GO enrichment analysis was conducted using the Singular Enrichment Analysis tool available on AgriGO (version 1.0) \citep{du2010} with a significance level of 5\% using Fisher statistical testing and Yekutieli multi-test adjustment. GO defines three distinct ontologies, called biological process, molecular function, and cellular component.
\subsection{Software availability}
The source code is available in a compressed archive in Supplemental File S2 and S3. It is also made available at \texttt{https://github.com/anjalisilva/MPLNClust} and is released under the open source MIT license.

\section{Results}

\subsection{Simulation Studies}
Simulation studies were conducted to illustrate the ability to recover the true underlying parameters for the mixtures of MPLN distribution. Two settings, one with two underlying clusters and another with three underlying clusters were considered. Table \ref{table_g2_3modelselection} summarizes the clustering results obtained via different model selection criteria and the corresponding average adjusted Rand index (ARI) values for two and three clusters settings. 

\begin{table*}[h!]
\centering
\caption{Number of clusters selected (and average ARI) for the comparative studies using different model selection criteria.}
\label{table_g2_3modelselection}
\begin{tabular}{lccccccccc}
\hline
& \multicolumn{4}{c}{$G=2$} & &\multicolumn{4}{c}{$G=3$}\\
& \multicolumn{4}{c}{\# of clusters selected (average ARI)} & &\multicolumn{4}{c}{\# of clusters selected (average ARI) }\\
\cline{2-5}\cline{7-10}
& BIC & ICL & AIC & AIC3  & & BIC & ICL & AIC & AIC3 \\
\hline\\
mix. of MPLN &  2  & 2  & 2  & 2 &&3  & 3 & 3  & 3  \\ 
&  (1.00) & (1.00) & (1.00) & (1.00) && (0.99) &  (0.99) &  (0.99) & (0.99) \\ \\
${\tt HTSCluster}$ & 3 & 3 & 3  & 3 && 4  & 4  & 4 & 4  \\ 
&  (-0.01) &  (-0.01) &  (-0.01) &  (-0.01) &&  (0.02) &  (0.02) &  (0.02) &  (0.02)\\ \\
{\tt Poisson.glm.mix}&3  & 3  & 3  & 3 && 4  & 4  & 4 & 4 \\
{\tt m = 1}& (0.09) &  (0.09) &  (0.09) &  (0.09)&&  (0.15) & (0.15) & (0.15) &  (0.15)\\ \\
{\tt Poisson.glm.mix}&3 & 3 & 3  & 3 & &4 & 4  & 4  & 4 \\
{\tt m = 2}& (0.00) &  (0.00) &  (0.00) &  (0.00)&& (0.04) & (0.04) & (0.04) &  (0.04)\\ \\
{\tt Poisson.glm.mix}&1-3 & 1-3  & 1-3  & 1-3 && 2-4& 2-4  & 2-4  & 2-4 \\
{\tt m = 3}& (0.00) & (0.00) &  (0.00) &  (0.00)&&  (0.02) & (0.02) &  (0.02) & (0.02)\\ \\
{\tt MBCluster.Seq}& 3  & 3  & 3 & 3 &&4  & 4  & 4  & 4\\
{\tt Poisson}&  (0.00) &  (0.00) &  (0.00) &  (0.00)&& (0.02) &  (0.02) &  (0.02) &  (0.02)\\ \\
{\tt MBCluster.Seq}&2 & 2  & 2  & 2&&2  & 2  & 2 & 2  \\
{\tt NB}&(-0.01) & (-0.01) &  (-0.01) & (-0.01)& &(0.00) & (0.00) &  (0.00) &  (0.00)\\
\hline
\end{tabular}
\end{table*}

The ARI values obtained are equal to or very close to one (see Table \ref{table_g2_3modelselection}). This is indicative that the algorithm for the mixtures of MPLN distribution is able to assign observations to the proper clusters, i.e., the clusters that were originally used to generate the simulation datasets. The parameter estimation results for $\boldsymbol{\mu}$ and $\boldsymbol{\Sigma}$ in the simulation study are summarized in Tables \ref{table_g2parameters} and \ref{table_g3parameters}, respectively for two and three clusters settings. 
\begin{table*}[h!]
\centering
\caption{Model parameters ($\boldsymbol{\vartheta}$) and the norms of the associated biases from the 50 runs for $G = 2$.}
\label{table_g2parameters}
\begin{tabular}{@{\extracolsep{-7pt}}lr@{}} \hline
True & $\| \textrm{Bias} \|$ \\ \\ \hline
$\boldsymbol{\mu}_1=(6.5,6,6,6,6,6)'$ &  0.11 \\
$\boldsymbol{\mu}_2=(2,2.5,2,2,2,2)'$ &  0.13 \\ 
$\boldsymbol{\Sigma}_1=\left[\begin{array}{cccccc}
1.24 & -0.36 & -0.51 & -0.04 & -0.54 & -0.39	\\
-0.36 & 1.30 & 0.11 & 0.23 & 0.90 & -0.77	\\
-0.51 &  0.11 & 1.25 & -0.44 & -0.01 & 0.04	\\
-0.04 & 0.23 & -0.44 & 1.09 & 0.84 & 0.38 \\
-0.54 & 0.90 & -0.01 & 0.84 & 1.41 & 0.21 \\
-0.39 & -0.77 & 0.04 & 0.38 & 0.21 & 1.33
\end{array}\right]$ & 0.05 \\
$\boldsymbol{\Sigma}_2=\left[\begin{array}{cccccc}
0.70 &  0.26  & -0.45  & -0.30  & -0.04 & -0.14 \\
0.26  & 0.70 & 0.19  & 0.27  & -0.07  & -0.05\\
-0.45 & 0.19  & 0.70  & 0.29  & 0.09  & 0.13\\
-0.30 & 0.27  & 0.29  & 0.70  & 0.25  & -0.04\\
-0.04 & -0.07  & 0.09  & 0.25  & 0.70  & 0.02\\
-0.14 & -0.05  & 0.13  & -0.04  & 0.02  & 0.70
\end{array}\right]$ & 0.15 \\
\hline
\end{tabular}
\end{table*}
\begin{table*}[h!]
\centering
\caption{Model parameters ($\boldsymbol{\vartheta}$) and the norms of the associated biases from the 30 runs for $G$ = 3.}
\label{table_g3parameters}
\begin{tabular}{@{\extracolsep{-7pt}}lr@{}}
\hline
True & $\| \textrm{Bias} \|$
\\
\hline
$\boldsymbol{\mu}_1=(3,3,3,3,3,3)'$ & 0.15 \\
$\boldsymbol{\mu}_2=(6.5,6.5,6.5,6.5,6,6.5)'$ &  0.13 \\
$\boldsymbol{\mu}_3= (1,-1,1,1,-1,1)'$ &  0.73\\
$\boldsymbol{\Sigma}_1=\left[\begin{array}{cccccc}
1.00 & -0.29 & -0.41& -0.04 & -0.41 & -0.31 \\
-0.29 & 1.00 & 0.08 & 0.19 & 0.66 & -0.59 \\
-0.41 &  0.08 &  1.00 & -0.38 & -0.01& 0.03 \\
-0.04 & 0.19 & -0.38 &  1.00 & 0.67 & 0.31\\
-0.41&  0.66 & -0.01 & 0.67 & 1.00 &  0.15\\
-0.31 & -0.59 & 0.03 &  0.31 &  0.15 & 1.00
\end{array}\right]$ &  1.55 \\
$\boldsymbol{\Sigma}_2=\left[\begin{array}{cccccc}
1.50 & -0.03 &  0.67 & 0.66 & -0.65 & -1.06 \\
-0.03 & 1.50 & -0.01 &  0.52 &  0.14 & -0.58 \\ 
 0.67 & -0.01 &  1.50 &  0.64 &  0.28 & -0.44 \\
 0.66 &  0.52 &  0.64 &  1.50 &  0.56 & -0.96 \\
-0.65 &  0.14 &  0.28 &  0.56 &  1.50 &  0.41 \\
-1.06 & -0.58 & -0.44 & -0.96 & 0.41 & 1.50
\end{array}\right]$ & 0.12\\
$\boldsymbol{\Sigma}_3=\left[\begin{array}{cccccc}
0.50 & 0.30 & -0.09 & -0.06 & 0.04 & -0.04 \\
0.30 & 0.50 & -0.02 & -0.02 & -0.07 & -0.17\\
-0.09 & -0.02 & 0.50 & 0.09 & 0.26 & 0.13\\
-0.06 & -0.02 & 0.09 & 0.50 &-0.01 & 0.19\\
 0.04 & -0.07 & 0.26 & -0.01 &  0.50 & -0.10\\
-0.04 & -0.17 & 0.13 & 0.19 & -0.10 & 0.50
\end{array}\right]$ &  0.20 \\ 
\hline
\end{tabular}
\end{table*}

Here, the norm of the bias of the model parameters are provided. Overall, the simulation experiments illustrate that the algorithm for mixtures of MPLN distribution is effective at parameter recovery.

\subsection{Comparison with Other Methods}
A number of mixture model-based clustering approaches for RNA-seq data exist \citep{omictool2014}. Three such methods available on the Comprehensive R Archive Network (CRAN) for {\sf R} include ${\tt HTSCluster}$ \citep{rau2011, rau2015}, ${\tt MBCluster.Seq}$ \citep{Si2014}, and ${\tt Poisson.glm.mix}$ \citep{papastamoulis2014}. Across all the methods, the discrete property of read counts is accommodated by clustering via Poisson mixtures with the exception of ${\tt MBCluster.Seq}$, which also offers clustering via NB mixtures. The ${\tt Poisson.glm.mix}$ offers three different parameterizations for the Poisson mean (by varying the regression coefficient), which will be termed ${\tt m = 1}$, ${\tt m = 2}$ and ${\tt m = 3}$ settings of ${\tt Poisson.glm.mix}$ \citep{papastamoulis2014}. All three methods were applied to the datasets generated using mixtures of MPLN distribution in the two and three cluster settings of the simulation study and the clustering results are summarized in Table \ref{table_g2_3modelselection}. All information criteria gave comparable clustering performance. It should be noted that for ${\tt Poisson.glm.mix}$ ${\tt m = 3}$, of the 50 datasets clustered under the two clusters setting, $G = 1$ was selected for 36 datasets, $G = 2$ was selected for 4 datasets, and $G = 3$ was selected for 10 datasets. Further analysis of the $G = 2$ models showed that for 3 out of the 4 datasets, all observations were assigned to one cluster making the other cluster an empty one. This resulted as the maximum \textit{a posteriori} probability occurred on one cluster for all observations. Similar results were also observed in the three clusters setting. Overall, the simulation results show that the highest cluster size considered in the range of cluster sizes was selected via model selection criteria for ${\tt HTSCluster}$, ${\tt Poisson.glm.mix}$ ${\tt m = 1}$, ${\tt Poisson.glm.mix}$ ${\tt m = 2}$, and ${\tt MBCluster.Seq}$ ${\tt Poisson}$. For ${\tt MBCluster.Seq}$ ${\tt NB}$, the lowest cluster size considered in the range of cluster sizes, $G = 2$, was selected in all simulation studies. Finally, for ${\tt Poisson.glm.mix m = 3}$, the choice seems  to vary among the entire range of cluster sizes considered.

\subsection{Application to transcriptome data}
To illustrate the applicability of the algorithm for mixtures of MPLN distribution, it is applied to a cranberry bean (\textit{Phaseolus vulgaris} L.) RNA-seq dataset generated in the study by Freixas Coutin et al. \citep{freixascoutin2017}. As a comparison, other clustering methods from the comparative study were also used. In this study RNA-seq was used to monitor transcriptional dynamics in the seed coats of darkening (D) and non-darkening (ND) cranberry bean recombinant inbred lines. Post harvest darkening of seed coat in certain classes of beans can be a significant problem, because this leads to a loss of economic value at marketplace \citep{Junk-Knievel2008}. Several factors can effect darkening of the seed coat, including the variety of bean in use, as some varieties darken faster than others \citep{Junk-Knievel2008}. Freixas Coutin et al. looked into D and ND cranberry beans at three developmental stages: early (E), intermediate (I) and mature (M). For each developmental stage, $3$ biological replicates were considered for a total of $18$ samples (see Table \ref{table_realdata}). The data are available on the NCBI Sequence Read Archive under the BioProject PRJNA380220.
\begin{table*}[h!]
\centering
\caption{Summary of RNA-seq dataset used for cluster analysis.}
\label{table_realdata}
\Fontvi 
\begin{tabular}{lllllllll} \hline
Organism & \begin{tabular}[c]{@{}l@{}} Year \\ \end{tabular} & Type & \begin{tabular}[c]{@{}l@{}}Number \\ of genes\end{tabular} & \begin{tabular}[c]{@{}l@{}}Replicates\\ per condition\end{tabular} & \begin{tabular}[c]{@{}l@{}}Read count \\ range\end{tabular} &  \begin{tabular}[c]{@{}l@{}} 5-95\% Read\\ count  range\end{tabular} & \begin{tabular}[c]{@{}l@{}}Library\\ size range\end{tabular} & \begin{tabular}[c]{@{}l@{}}Platform and\\ Instrument\end{tabular} \\ \\\hline \\
\textit{P. vulgaris} & 2017 & RNA  & 1336                                                                                                      & (3,3,3,3,3,3)  & (0--483,965) & (205--3652)         & (937,559--1,870,947)                                                                                                    & \begin{tabular}[c]{@{}l@{}}Illumina \\ HiSeq 2000\end{tabular} \\  \\\hline                                               
\end{tabular}
\end{table*}

The clustering results are summarized in Table 5. For the mixtures of MPLN distribution, all information criteria selected a model with $G = 4$, with the exception of AIC, which selected a $G = 5$ model. Recall that AIC is known to favor the complex model with more parameters. For the $G = 4$ model, there were 71, 731, 415 and 119 genes in each respective cluster. For the $G = 5$ model, each cluster contained 113,  22, 230, 342 and 629 genes. A comparison of these two models revealed that each cluster from the $G = 4$ model was further divided into separate clusters in the $G = 5$ model. As per agreement among three information criteria, only the $G = 4$ model is used for downstream analysis. For ${\tt MBCluster.Seq}$ ${\tt NB}$, a model with $G = 2$ was selected. Here, cluster~1 contained 467 genes and cluster~2 contained 869 genes. A comparison of this model with that of $G = 4$, from mixtures of MPLN distribution, did not reveal any significant patterns. For all other clustering methods, information criteria selected a model with $G = 11$, the maximum number of clusters considered in the range of clusters. The composition of clusters differed across methods (see Supplemental File S1), with the smallest cluster having 39 genes and largest cluster having 274 genes (both resulted from ${\tt MBCluster.Seq}$ ${\tt Poisson}$). 
\begin{table*}[h!]
\label{table_realdata_modelselection}
\centering
\caption{Number of clusters selected using different model selection criteria for the RNA-seq data.}
\begin{tabular}{@{\extracolsep{3pt}}lrrrr@{}}
\hline
 & \multicolumn{4}{c}{Model selection criteria}\\
\cline{2-5} 
& BIC & ICL & AIC & AIC3\\\hline
mixtures of MPLN & 4  & 4 & 5 &  4  \\
${\tt HTSCluster}$ & 11  & 11 & 11 &  11\\
\begin{tabular}[c]{@{}l@{}}${\tt Poisson.glm.mix}, {\tt m = 1}$ \end{tabular} & 11  & 11 & 11 &  11   \\
\begin{tabular}[c]{@{}l@{}}${\tt Poisson.glm.mix}, {\tt m = 2}$ \end{tabular} & 11  & 11 & 11 &  11  \\
\begin{tabular}[c]{@{}l@{}}${\tt Poisson.glm.mix}, {\tt m = 3}$ \end{tabular} & 11  & 11 & 11 &  11 \\
\begin{tabular}[c]{@{}l@{}}${\tt MBCluster.Seq}, {\tt Poisson}$ \end{tabular}  & 11  & 11 & 11 &  11  \\
\begin{tabular}[c]{@{}l@{}}${\tt MBCluster.Seq}, {\tt NB}$ \end{tabular}   & 2  & 2 & 2 &  2\\\hline
\end{tabular}
\end{table*}
 
Downstream analysis of $G = 4$ model of mixtures of MPLN distribution, using Gene Ontology (GO), revealed 3 GO terms which exhibited enrichment for cluster~1 genes based on significant p-values. These belonged pathogenesis (GO:0009405), multi-organism process (GO:0051704) and nutrient reservoir activity (GO:0045735). No GO terms exhibited enrichment for cluster~2 genes and 13 GO terms exhibited enrichment for cluster~3 genes. Of these 13, one belonged to biological processes, oxidation reduction (GO:0055114). All others belonged to molecular function and included oxidoreductase activity (GO:0016491, GO:0016705, GO:0016614, GO:0016616, GO:0016706), enzyme activity (GO:0004857, GO:0050662),  binding (GO:0005506, GO:0048037, GO:0046906, GO:0020037) and dehydrogenase activity (GO:0008762). For cluster~4, 20 GO terms exhibited enrichment, of which 15 belonged to biological processes and 5 belonged to molecular functions. These included GO terms associated with lipid metabolic or related processes (GO:0006631, GO:0006629, GO:0032787, GO:0019752, GO:0043436, GO:0042180, GO:0006082, GO:0044255, GO:0044281), lipid biosynthesis or related process (GO:0008610, GO:0006633, GO:0044283, GO:0046394, GO:0016053, GO:0009058), transferase activity (GO:0016747, GO:0016746, GO:0008415) and synthase activity (GO:0004312, GO:0004315). Upon further examination, it was identified that 32 genes in this cluster are annotated as flavonoid/ proanthocyanidin biosynthesis genes in the \textit{P. vulgaris} genome. For the ${\tt MBCluster.Seq}$ ${\tt NB}$ model, no significant GO terms were found for cluster~1. For cluster~2, 36 GO terms were identified, 24 of which belonged to biological processes and 12 of which belonged to molecular functions. These included metabolic processes (GO:0006629, GO:0044281, GO:0043436, GO:0006082, GO:0008202, GO:0019752, GO:0042180, GO:0009072, GO:0046417, GO:0006631, GO:0006519, GO:0032787), biosynthetic processes (GO:0008610, GO:0006694, GO:0044283, GO:0006633, GO:0016053, GO:0046394, GO:0009073), oxidation reduction (GO:0055114, GO:0016616, GO:0016614, GO:0016491), pathogenesis (GO:0009405), drug transport (GO:0006855, GO:0015893, GO:0015238), response to drug (GO:0042493), synthase activity (GO:0004315, GO:0004312), dehydrogenase activity (GO:0003854, GO:0033764, GO:0016229), binding (GO:0050662, GO:0048037) and catalytic activity (GO:0003824). 

Overall, the transcriptome data analysis shows that the highest cluster size considered in the range of cluster sizes was selected using information criteria for ${\tt HTSCluster}$, all settings of ${\tt Poisson.glm.mix}$ and ${\tt MBCluster.Seq}$ ${\tt Poisson}$. For ${\tt MBCluster.Seq}$ ${\tt NB}$, the lowest cluster size considered in the range of cluster sizes, $G = 2$, was selected. However, for mixtures of MPLN distribution, the selection was in between the range. 

\section{Discussion}\label{sec:discussion}
A model-based clustering technique for RNA-seq data was introduced. The approach utilizes a mixture of MPLN distributions, which has not previously been used for model-based clustering of RNA-seq data. 
The MPLN distribution is able to describe a wide range of correlation and overdispersion situations, and is ideal for modeling RNA-seq data, which is generally overdispersed and can be negatively correlated. Importantly, the hidden layer of the MPLN distribution is a multivariate Gaussian distribution which accounts for the covariance structure of the data. As a result, independence does not need to be assumed between variables in clustering applications. The Gaussian mixture models are the most popular models and have shown to be a powerful tool for many applications \citep{yeung2001, zhong2003}. Extensions applicable to Gaussian  mixture models, such as subspace clustering techniques, will be applicable to the MPLN distribution at the hidden layer level \citep{mcnicholas2008, bouveyron2012}.

The MCMC-EM algorithm is employed for parameter estimation in the mixtures of MPLN distribution. MCMC algorithms have become popular due to their ability to fit complex models where the standard methods are difficult to apply. However, determining convergence of MCMC can be difficult as it is hard to determine if the stationary distribution obtained via MCMC is same as the posterior distribution of interest. Convergence criteria proposed to date include checking if the sampled distribution has reached stationarity or comparing if the distributions obtained from MCMC for different runs show negligible differences. A combination of these diagnostics, the potential scale reduction factor and effective number of samples, are used to assess the convergence of MCMC chains. The Heidelberger and Welch's convergence diagnostic is used to test if the sampled values come from a stationary distribution.

For this analysis, the MCMC-EM algorithm was set to start with a modest number of ${\tt RStan}$ iterations and was increased, for each MCMC-EM iteration, as the algorithm proceed. Intuitively, the MCMC-EM algorithm should be started with a modest Monte Carlo sample size, when the parameter values are relatively far from the MLE. During this time, the MCMC-EM update makes a substantial jump and less precision is required for the Monte Carlo approximation to that jump \citep{neath2012}. As the algorithm proceeds, the parameter values should get close to the MLE and the EM update will be small, such that greater precision is required for the Monte Carlo approximation \citep{neath2012}. 

Using simulation studies, we were able to illustrate that our algorithm is effective at parameter recovery and returned favorable clustering results. During the comparison study, it was observed that other methods from the current literature failed to identify the true number of underlying clusters majority of the time. Occasionally, when the correct number of underlying clusters were identified, misclassification and empty clusters were present. Low ARI values ($<0.2$) were observed for all other methods. Conclusions from these simulations should be drawn with some caution as the data were simulated via mixtures of MPLN distribution. The simulation of RNA-seq data in the context of clustering applications remains an open question \citep{rau2015}. This topic has been investigated in the context of differential expression analysis and various methods available for this purpose can be found in OMICtools \citep{omictool2014}. Further, empty clusters were observed, which resulted when MAP classification occurred on fewer clusters than the number of subgroups considered in the clustering run. A check should be inserted to compare whether the final resulting number of clusters via MAP and the number of subgroups are the same. 

The transcriptome data analysis showed the applicability of mixtures of MPLN distribution on RNA-seq data. Cluster analysis of differentially expressed genes identified in the study by Freixas Coutin et al. resulted a four cluster model via the mixtures of MPLN distribution. GO analysis revealed that majority of the genes in cluster~1 were involved in multi-organism process, which is physiological interaction between organisms. Genes in cluster~3 were associated with oxidoreductase activity and cluster~4 was found to contain genes involved in lipid metabolic and biosynthesis processes. Further analysis revealed that this cluster also contains genes that were annotated as flavonoid/proanthocyanidin biosynthesis. In the analysis of transcriptome data using other methods, it was noted that information criteria tend to select the highest cluster size considered in the range of clusters for ${\tt HTSCluster}$, ${\tt Poisson.glm.mix}$ and ${\tt MBCluster.Seq}$ ${\tt Poisson}$. However, for ${\tt MBCluster.Seq}$ ${\tt NB}$, the lowest cluster size considered in the range of clusters, $G = 2$, is selected. Similar observations were also made during the comparison analysis using simulated data. This could potentially imply that models assuming independence between the variables are not providing a good fit to the data. As a result, under-fitting or over-fitting of the number of clusters may have occurred. However, when using mixtures of MPLN, in simulation studies, the model selection criteria were able to select the correct number of components.

A direction for future work would be to investigate subspace clustering methods to overcome \textit{curse of dimensionality} as high-dimensional RNA-seq datasets become frequent. Subspace clustering involves clustering data in low-dimensional subspaces without reducing the dimension, as dimension reduction can imply an information loss for discriminating the groups \citep{bouveyron2012}. In this type of clustering, restrictions are introduced to mixture parameters with the aim of obtaining parsimonious models, which are sufficiently flexible for clustering purposes.

\subsection*{Acknowledgments}
{\small This work was supported by Discovery Grant from the Natural Sciences and Engineering Research Council of Canada (Subedi), Queen Elizabeth II Aiming for the Top Scholarship (Silva), and Arthur Richmond Memorial Scholarship (Silva). The authors acknowledge the computational support provided by Dr.\ Marcelo Ponce at the SciNet HPC Consortium, University of Toronto, ON, Canada.}


{\small
\begin{thebibliography}{}

\bibitem[\protect\citeauthoryear{Aitchison and Ho}{Aitchison and
  Ho}{1989}]{aitchison1989}
Aitchison, J. and C.~H. Ho (1989).
\newblock The multivariate {P}oisson-log normal distribution.
\newblock {\em Biometrika\/}~{\em 76\/}(4), 643--653.

\bibitem[\protect\citeauthoryear{Akaike}{Akaike}{1973}]{akaike1973}
Akaike, H. (1973).
\newblock Information theory and an extension of the maximum likelihood
  principle.
\newblock In {\em Second International Symposium on Information Theory}, pp.\
  267--281. Springer Verlag.

\bibitem[\protect\citeauthoryear{Anders and Huber}{Anders and
  Huber}{2010}]{anders2010}
Anders, S. and W.~Huber (2010).
\newblock Differential expression analysis for sequence count data.
\newblock {\em Genome Biol\/}~{\em 11}, 1--12.

\bibitem[\protect\citeauthoryear{Anders, Pyl, and Huber}{Anders
  et~al.}{2015}]{anders2015}
Anders, S., P.~T. Pyl, and W.~Huber (2015).
\newblock {HTSeq-a Python framework to work with high-throughput sequencing
  data}.
\newblock {\em Bioinformatics\/}~{\em 31}, 166--169.

\bibitem[\protect\citeauthoryear{Annis, Miller, and Palmeri}{Annis
  et~al.}{2016}]{stan}
Annis, J., B.~J. Miller, and T.~J. Palmeri (2016).
\newblock {Bayesian inference with Stan: A tutorial on adding custom
  distributions}.
\newblock {\em Behav Res Methods\/}~{\em 49}, 1--24.

\bibitem[\protect\citeauthoryear{Biernacki, Celeux, and Govaert}{Biernacki
  et~al.}{2000}]{biernacki2000}
Biernacki, C., G.~Celeux, and G.~Govaert (2000).
\newblock Assessing a mixture model for clustering with the integrated
  classification likelihood.
\newblock {\em IEEE Transactions on Pattern Analysis and Machine
  Intelligence\/}~{\em 22\/}(7), 719--725.

\bibitem[\protect\citeauthoryear{Booth and Hobert}{Booth and
  Hobert}{1999}]{hobert1999}
Booth, J.~G. and J.~P. Hobert (1999).
\newblock Maximizing generalized linear mixed model likelihoods with an
  automated {M}onte {C}arlo {EM} algorithm.
\newblock {\em Journal of the Royal Stat Society: Series~B\/}~{\em 61},
  265--285.

\bibitem[\protect\citeauthoryear{Bouveyron and Brunet}{Bouveyron and
  Brunet}{2013}]{bouveyron2012}
Bouveyron, C. and C.~Brunet (2013).
\newblock Model-based clustering of high-dimensional data: A review.
\newblock {\em Computational Statistics and Data Analysis\/}~{\em 71}, 52--78.

\bibitem[\protect\citeauthoryear{Bozdogan}{Bozdogan}{1994}]{Bozdogan1994}
Bozdogan, H. (1994).
\newblock Mixture-model cluster analysis using model selection criteria and a
  new informational measure of complexity.
\newblock In {\em Proceedings of the First US/Japan Conference on the Frontiers
  of Statistical Modeling: An Informational Approach: Volume 2 Multivariate
  Statistical Modeling}, pp.\  69--113. Dordrecht: Springer Netherlands.

\bibitem[\protect\citeauthoryear{Cowles and Carlin}{Cowles and
  Carlin}{1996}]{cowles1996}
Cowles, M.~K. and B.~P. Carlin (1996).
\newblock Markov chain {M}onte {C}arlo convergence diagnostics: a comparative
  review.
\newblock {\em Journal of the American Statistical Association\/}~{\em
  91\/}(434), 883--904.

\bibitem[\protect\citeauthoryear{Dempster, Laird, and Rubin}{Dempster
  et~al.}{1977}]{dempster1977}
Dempster, A.~P., N.~M. Laird, and D.~B. Rubin (1977).
\newblock Maximum likelihood from incomplete data via the {EM} algorithm.
\newblock {\em Journal of the Royal Statistical Society: Series~B\/}~{\em 39},
  1--38.

\bibitem[\protect\citeauthoryear{D'haeseleer}{D'haeseleer}{2005}]{dhaeseleer2005}
D'haeseleer, P. (2005).
\newblock How does gene expression clustering work?
\newblock {\em Nat Biotechnol\/}~{\em 23}, 1499--1501.

\bibitem[\protect\citeauthoryear{Du, Zhou, Ling, Zhang, and Su}{Du
  et~al.}{2010}]{du2010}
Du, Z., X.~Zhou, Y.~Ling, Z.~Zhang, and Z.~Su (2010).
\newblock {agriGO: a GO analysis toolkit for the agricultural community}.
\newblock {\em Nucleic Acids Res\/}~{\em 38}, W64--W70.

\bibitem[\protect\citeauthoryear{Esnaola, Puig, Gonzalez, Castelo, and
  Gonzalez}{Esnaola et~al.}{2013}]{esnaola2013}
Esnaola, M., P.~Puig, D.~Gonzalez, R.~Castelo, and J.~R. Gonzalez (2013).
\newblock A flexible count data model to fit the wide diversity of expression
  profiles arising from extensively replicated {RNA}-seq experiments.
\newblock {\em BMC Bioinformatics\/}~{\em 14\/}(254).

\bibitem[\protect\citeauthoryear{Freixas-Coutin, Munholland, Silva, Subedi,
  Lukens, Crosby, Pauls, and Bozzo}{Freixas-Coutin
  et~al.}{2017}]{freixascoutin2017}
Freixas-Coutin, J.~A., S.~Munholland, A.~Silva, S.~Subedi, L.~Lukens, W.~L.
  Crosby, K.~P. Pauls, and G.~G. Bozzo (2017).
\newblock Proanthocyanidin accumulation and transcriptional responses in the
  seed coat of cranberry beans {(\textit{Phaseolus vulgaris} L)} with different
  susceptibility to postharvest darkening.
\newblock {\em BMC Plant Biol\/}~{\em 17\/}(89).

\bibitem[\protect\citeauthoryear{Gelman, Carlin, Stern, Dunson, Vehtari, and
  Rubin}{Gelman et~al.}{2013}]{gelman2013}
Gelman, A., J.~B. Carlin, H.~S. Stern, D.~B. Dunson, A.~Vehtari, and D.~B.
  Rubin (2013).
\newblock {\em Bayesian Data Analysis}.
\newblock Boca Raton, FL: Chapman \& Hall/CRC Press.

\bibitem[\protect\citeauthoryear{Gelman and Rubin}{Gelman and
  Rubin}{1992}]{gelman1992}
Gelman, A. and D.~B. Rubin (1992).
\newblock Inference from iterative simulation using multiple sequences.
\newblock {\em Stat Sci\/}~{\em 7}, 457--472.

\bibitem[\protect\citeauthoryear{Georgescu, Desassis, Soubeyrand, Kretzschmar,
  and Senoussi}{Georgescu et~al.}{2011}]{vera2011}
Georgescu, V., N.~Desassis, S.~Soubeyrand, A.~Kretzschmar, and R.~Senoussi
  (2011).
\newblock A hierarchical model for multivariate data of different types and
  maximum likelihood estimation.
\newblock Technical Report RR-46, {INRIA}, Saclay, Ile-de-France.

\bibitem[\protect\citeauthoryear{Heidelberger and Welch}{Heidelberger and
  Welch}{1983}]{heidelberger1983}
Heidelberger, P. and P.~D. Welch (1983).
\newblock Simulation run length control in the presence of an initial
  transient.
\newblock {\em Operations Research\/}~{\em 31}, 1109--1144.

\bibitem[\protect\citeauthoryear{Henry, Bandrowski, Pepin, Gonzalez, and
  Desfeux}{Henry et~al.}{2014}]{omictool2014}
Henry, V.~J., A.~E. Bandrowski, A.~Pepin, B.~J. Gonzalez, and A.~Desfeux
  (2014).
\newblock {OMIC}tools: an informative directory for multi-omic data analysis.
\newblock {\em Database (Oxford)\/}~{\em 2014}.

\bibitem[\protect\citeauthoryear{Hoffman and Gelman}{Hoffman and
  Gelman}{2014}]{nuts2014}
Hoffman, M.~D. and A.~Gelman (2014).
\newblock The no-u-turn sampler: Adaptively setting path lengths in
  {H}amiltonian {M}onte {C}arlo.
\newblock {\em Journal of Machine Learning Research\/}~{\em 15}, 1593--1623.

\bibitem[\protect\citeauthoryear{Hubert and Arabie}{Hubert and
  Arabie}{1985}]{hubert1985}
Hubert, L. and P.~Arabie (1985).
\newblock Comparing partitions.
\newblock {\em Journal of Classification\/}~{\em 2}, 193--218.

\bibitem[\protect\citeauthoryear{Junk-Knievel, Vandenberg, and
  Bett}{Junk-Knievel et~al.}{2008}]{Junk-Knievel2008}
Junk-Knievel, D.~C., A.~Vandenberg, and K.~E. Bett (2008).
\newblock Slow darkening in {pinto bean (\textit{Phaseolus vulgaris} L)} seed
  coats is controlled by a single major gene.
\newblock {\em Crop Science\/}~{\em 48}, 189--193.

\bibitem[\protect\citeauthoryear{Katz}{Katz}{1981}]{katz1981}
Katz, R.~W. (1981).
\newblock On some criteria for estimating the order of a {M}arkov chain.
\newblock {\em Technometrics\/}~{\em 23\/}(3), 243--249.

\bibitem[\protect\citeauthoryear{Kvam, Liu, and Si}{Kvam
  et~al.}{2012}]{kvam2012}
Kvam, V.~M., P.~Liu, and Y.~Si (2012).
\newblock A comparison of statistical methods for detecting differentially
  expressed genes from {RNA-seq} data.
\newblock {\em Am J Bot\/}~{\em 99}, 248--256.

\bibitem[\protect\citeauthoryear{Li, Handsaker, Wysoker, Fennell, Ruan, Homer,
  Marth, Abecasis, Durbin, and Subgroup}{Li et~al.}{2009}]{li2009}
Li, H., B.~Handsaker, A.~Wysoker, T.~Fennell, J.~Ruan, N.~Homer, G.~Marth,
  G.~Abecasis, R.~Durbin, and .~G. P. D.~P. Subgroup (2009).
\newblock {The Sequence alignment/map (SAM) format and SAMtools}.
\newblock {\em Bioinformatics\/}~{\em 25}, 2078--2079.

\bibitem[\protect\citeauthoryear{McCarthy, Chen, and Smyth}{McCarthy
  et~al.}{2012}]{mccarthy2012}
McCarthy, J.~D., Y.~Chen, and K.~G. Smyth (2012).
\newblock Differential expression analysis of multifactor {RNA-Seq} experiments
  with respect to biological variation.
\newblock {\em Nucleic Acids Res\/}~{\em 40}, 4288--4297.

\bibitem[\protect\citeauthoryear{McLachlan and Basford}{McLachlan and
  Basford}{1988}]{mclachlanandbasford1998}
McLachlan, G.~J. and K.~E. Basford (1988).
\newblock {\em Mixture Models Inference and Applications to Clustering}.
\newblock New York: Marcel Dekker.

\bibitem[\protect\citeauthoryear{McLachlan and Peel}{McLachlan and
  Peel}{2000}]{mclachlan2000}
McLachlan, G.~J. and D.~Peel (2000).
\newblock {\em {Finite Mixture Models}}.
\newblock New York: Wiley.

\bibitem[\protect\citeauthoryear{McNicholas}{McNicholas}{2016}]{mcnicholas2016}
McNicholas, P.~D. (2016).
\newblock {\em {Mixture Model-based Classification}}.
\newblock Boca Raton: Chapman and Hall/CRC Press.

\bibitem[\protect\citeauthoryear{McNicholas and Murphy}{McNicholas and
  Murphy}{2008}]{mcnicholas2008}
McNicholas, P.~D. and T.~B. Murphy (2008).
\newblock {Parsimonious Gaussian mixture models}.
\newblock {\em Statistics and Computing\/}~{\em 18\/}(3), 285--296.

\bibitem[\protect\citeauthoryear{McNicholas, Murphy, McDaid, and
  Frost}{McNicholas et~al.}{2010}]{mcnicholas2010parallel}
McNicholas, P.~D., T.~B. Murphy, A.~F. McDaid, and D.~Frost (2010).
\newblock Serial and parallel implementations of model-based clustering via
  parsimonious {G}aussian mixture models.
\newblock {\em Computational Statistics and Data Analysis\/}~{\em 54\/}(3),
  711--723.

\bibitem[\protect\citeauthoryear{Neath}{Neath}{2012}]{neath2012}
Neath, R.~C. (2012).
\newblock On convergence properties of the {M}onte {C}arlo {EM} algorithm.
\newblock arXiv preprint arXiv:1206.4768.

\bibitem[\protect\citeauthoryear{Oshlack, Robinson, and Young}{Oshlack
  et~al.}{2010}]{oshlack2010}
Oshlack, A., M.~D. Robinson, and M.~D. Young (2010).
\newblock From {RNA-seq} reads to differential expression results.
\newblock {\em Genome Biol\/}~{\em 11}, doi: 10.1186/gb--2010--11--12--220.

\bibitem[\protect\citeauthoryear{Papastamoulis, Martin-Magniette, and
  Maugis-Rabusseau}{Papastamoulis et~al.}{2014}]{papastamoulis2014}
Papastamoulis, P., M.~Martin-Magniette, and C.~Maugis-Rabusseau (2014).
\newblock On the estimation of mixtures of {P}oisson regression models with
  large number of components.
\newblock {\em Computational Statistics and Data Analysis\/}~{\em 93}, 97--106.

\bibitem[\protect\citeauthoryear{Plummer, Best, Cowles, and Vines}{Plummer
  et~al.}{2006}]{coda}
Plummer, M., N.~Best, K.~Cowles, and K.~Vines (2006).
\newblock {CODA}: Convergence diagnosis and output analysis for {MCMC}.
\newblock {\em R News\/}~{\em 6\/}(1), 7--11.

\bibitem[\protect\citeauthoryear{Qiu and Joe}{Qiu and
  Joe}{2015}]{clusterGeneration2015}
Qiu, W. and H.~Joe (2015).
\newblock {\em {clusterGeneration}: Random Cluster Generation (with Specified
  Degree of Separation)}.
\newblock R package version 1.3.4.

\bibitem[\protect\citeauthoryear{{R Core Team}}{{R Core Team}}{2017}]{r2014}
{R Core Team} (2017).
\newblock {\em R: A Language and Environment for Statistical Computing}.
\newblock Vienna, Austria: R Foundation for Statistical Computing.

\bibitem[\protect\citeauthoryear{Rau, Celeux, Martin-Magniette, and
  Maugis-Rabusseau}{Rau et~al.}{2011}]{rau2011}
Rau, A., G.~Celeux, M.~Martin-Magniette, and C.~Maugis-Rabusseau (2011).
\newblock Clustering high-throughput sequencing data with {P}oisson mixture
  models.
\newblock Technical Report RR-7786, {INRIA}, Saclay, Ile-de-France.

\bibitem[\protect\citeauthoryear{Rau, Maugis-Rabusseau, Martin-Magniette, and
  Celeux}{Rau et~al.}{2015}]{rau2015}
Rau, A., C.~Maugis-Rabusseau, M.~L. Martin-Magniette, and G.~Celeux (2015).
\newblock Co-expression analysis of high-throughput transcriptome sequencing
  data with {P}oisson mixture models.
\newblock {\em Bioinformatics\/}~{\em 31\/}(9), 1420--1427.

\bibitem[\protect\citeauthoryear{{Revolution Analytics} and Weston}{{Revolution
  Analytics} and Weston}{2015}]{foreach2015}
{Revolution Analytics} and S.~Weston (2015).
\newblock {\em {foreach}: Provides Foreach Looping Construct for {R}}.
\newblock R package version 1.4.3.

\bibitem[\protect\citeauthoryear{Roberts, Pimentel, Trapnell, and
  Pachter}{Roberts et~al.}{2011}]{roberts2011}
Roberts, A., H.~Pimentel, C.~Trapnell, and L.~Pachter (2011).
\newblock Identification of novel transcripts in annotated genomes using
  {RNA-Seq}.
\newblock {\em Bioinformatics\/}~{\em 27}, 2325--2329.

\bibitem[\protect\citeauthoryear{Robinson, McCarthy, and Smyth}{Robinson
  et~al.}{2010}]{edgeR2010}
Robinson, M.~D., D.~J. McCarthy, and G.~K. Smyth (2010).
\newblock {edgeR}: a bioconductor package for differential expression analysis
  of digital gene expression data.
\newblock {\em Bioinformatics\/}~{\em 26}, 139--140.

\bibitem[\protect\citeauthoryear{Robinson and Oshlack}{Robinson and
  Oshlack}{2010}]{robinson2010}
Robinson, M.~D. and A.~Oshlack (2010).
\newblock A scaling normalization method for differential expression analysis
  of {RNA}-seq data.
\newblock {\em Genome Biol\/}~{\em 11\/}(R25).

\bibitem[\protect\citeauthoryear{{RStudio Team}}{{RStudio
  Team}}{2016}]{rstudio2012}
{RStudio Team} (2016).
\newblock {\em {RS}tudio: Integrated Development Environment for {R}}.
\newblock Boston: RStudio, Inc.

\bibitem[\protect\citeauthoryear{Schwarz}{Schwarz}{1978}]{schwarz1978}
Schwarz, G. (1978).
\newblock Estimating the dimension of a model.
\newblock {\em The Annals of Statistics\/}~{\em 6}, 461--464.

\bibitem[\protect\citeauthoryear{Shibata}{Shibata}{1976}]{shibata1976}
Shibata, R. (1976).
\newblock Selection of the order of an autoregressive model by {A}kaike's
  information criterion.
\newblock {\em Biometrika\/}~{\em 63\/}(1), 117--126.

\bibitem[\protect\citeauthoryear{Si, Liu, Li, and Brutnell}{Si
  et~al.}{2014}]{Si2014}
Si, Y., P.~Liu, P.~Li, and T.~P. Brutnell (2014).
\newblock Model-based clustering for {RNA}-seq data.
\newblock {\em Bioinformatics\/}~{\em 30}, 197--205.

\bibitem[\protect\citeauthoryear{{Stan Development Team}}{{Stan Development
  Team}}{2016}]{rstan}
{Stan Development Team} (2016).
\newblock {RStan}: the {R} interface to {Stan}.
\newblock R package version 2.15.1.

\bibitem[\protect\citeauthoryear{Tunaru}{Tunaru}{2002}]{tunaru2002}
Tunaru, R. (2002).
\newblock Hierarchical {B}ayesian models for multiple count data.
\newblock {\em Austrian J Stat\/}~{\em 31}, 221--229.

\bibitem[\protect\citeauthoryear{Wang, Gerstein, and Snyder}{Wang
  et~al.}{2009}]{wang2009}
Wang, Z., M.~Gerstein, and M.~Snyder (2009).
\newblock {RNA-Seq}: a revolutionary tool for transcriptomics.
\newblock {\em Nat Rev Genet\/}~{\em 10}, 57--63.

\bibitem[\protect\citeauthoryear{Wei and Tanner}{Wei and
  Tanner}{1990}]{wei1990a}
Wei, G. C.~G. and M.~A. Tanner (1990).
\newblock A {M}onte {C}arlo implementation of the {EM} algorithm and the {P}oor
  {M}an's data augmentation algorithms.
\newblock {\em Journal of the American Statistical Association\/}~{\em 85},
  699--704.

\bibitem[\protect\citeauthoryear{Wolfe}{Wolfe}{1965}]{wolfe1965}
Wolfe, J.~H. (1965).
\newblock A computer program for the maximum likelihood analysis of types.
\newblock Technical Bulletin 65--15, US\ Naval Personnel Research Activity.

\bibitem[\protect\citeauthoryear{Yeung, Fraley, Murua, Raftery, and
  Ruzzo}{Yeung et~al.}{2001}]{yeung2001}
Yeung, K.~Y., C.~Fraley, A.~Murua, A.~E. Raftery, and W.~L. Ruzzo (2001).
\newblock Model-based clustering and data transformations for gene expression
  data.
\newblock {\em Bioinformatics\/}~{\em 17}, 977--987.

\bibitem[\protect\citeauthoryear{Zhong and Ghosh}{Zhong and
  Ghosh}{2003}]{zhong2003}
Zhong, S. and J.~Ghosh (2003).
\newblock A unified framework for model-based clustering.
\newblock {\em Journal of Machine Learning Research\/}~{\em 4}, 1001--1037.

\end{thebibliography}
}
\end{document}